\documentclass{styles/svproc}
\usepackage{url}

\usepackage{lineno}
\usepackage{graphicx}
\usepackage{multicol}
\usepackage{footmisc}
\usepackage{subfig}

\begin{document}
\mainmatter              
\title{Strangeness production with respect to high momentum hadrons in pp and p-Pb collisions with ALICE at the LHC}
\titlerunning{Strangeness Production in pp and p-Pb Collisions}  
%
\author{Justin T. Blair for the ALICE Collaboration}
\authorrunning{Justin T. Blair} 
%
%
\institute{University of Texas at Austin, Austin TX 78712\\
\email{jtblair@utexas.edu}}

\maketitle              

\begin{abstract}
In order to understand strangeness production mechanisms, one can study the correlations of hadrons with hidden strangeness (e.g. the $\phi$ meson) and open strangeness ($\mathrm{K}^{0}_{\mathrm{S}}$, $\Lambda$ and $\bar{\Lambda}$) in hard (jet) processes and in soft (bulk) processes. Two-particle angular correlations with $\mathrm{K}^{0}_{\mathrm{S}}$ triggers allow for the study of fragmentation in pp collisions as a function of multiplicity. Similarly, correlations with associated $\phi$ mesons in p-Pb collisions allow both the jet and the underlying event components of strange particle production to be probed. Presented are $\mathrm{K}^{0}_{\mathrm{S}}$-h correlations measured by ALICE \cite{ALICE} in pp collisions at $\sqrt{s} = 13$ TeV, and h-$\phi$ correlations measured in p-Pb collisions at $\sqrt{s_{\mathrm{NN}}} = 5.02$ TeV.

\end{abstract}
\section{Introduction}	

Historically, an enhanced production of multi-strange hadrons in heavy-ion collisions was seen as a clear signature of the presence of a QGP \cite{Rafelski:1982pu}. Recent results have shown that this increase in multi-strange hadrons is also present in high multiplicity pp and p-A collisions with a smooth increase in strange hadron production as a function of charged particle multiplicity \cite{ppenhance,Adam:2015vsf,Adam:2016bpr}. This increase is seen across all collision systems, and is more prominent for multi-strange hadrons.

In order to probe the origin of this strangeness enhancement in small systems, it is necessary to study the different production mechanisms of strange quarks. Jet production can be separated from production in the underlying event by measuring angular correlations \cite{ALICE:2015pPb}, allowing for the study of strangeness production in both hard and soft scatterings. Using a high momentum strange hadron ($\mathrm{K}^{0}_{\mathrm{S}}$) as a jet proxy, the fragmentation of jets containing strange quarks can be studied as a function of multiplicity. Similarly, measuring associated strange hadrons ($\phi$) correlated to an high momentum trigger hadron can separate strange quarks produced in hard processes (jet-like) from those produced in soft processes (underlying event).

\section{$\mathrm{K}^{0}_{\mathrm{S}}$-h Angular Correlations}

To study jets containing strangeness, high momentum $\mathrm{K}^{0}_{\mathrm{S}}$ triggers are selected as a jet proxy in pp collisions at $\sqrt{s} = 13$ TeV. The trigger $\mathrm{K}^{0}_{\mathrm{S}}$ are reconstructed from the $\mathrm{K}^{0}_{\mathrm{S}} \rightarrow \pi^{+}\pi^{-}$ decay channel (B.R $\approx 69\%$). These decay pions can be identified in the ALICE detector using particle identification (PID) cuts on signals from both the Time Projection Chamber (TPC) and Time of Flight (TOF) detectors, as well as topological cuts to account for the displaced decay vertex of the $\mathrm{K}^{0}_{\mathrm{S}}$. Event multiplicity is determined using the V0 detectors at large forward and backward rapidity (V0M estimator). Yields of associated hadrons are measured in both the near-side and away-side jet peaks for different event multiplicities: 0-10\% (highest multiplicity), 10-50\%, 50-100\%, and 0-100\% inclusive. Results are compared to predictions from simulated PYTHIA \cite{Sjostrand:2014zea} events.

\begin{figure}
\centering
	\includegraphics[width = 3.2in]{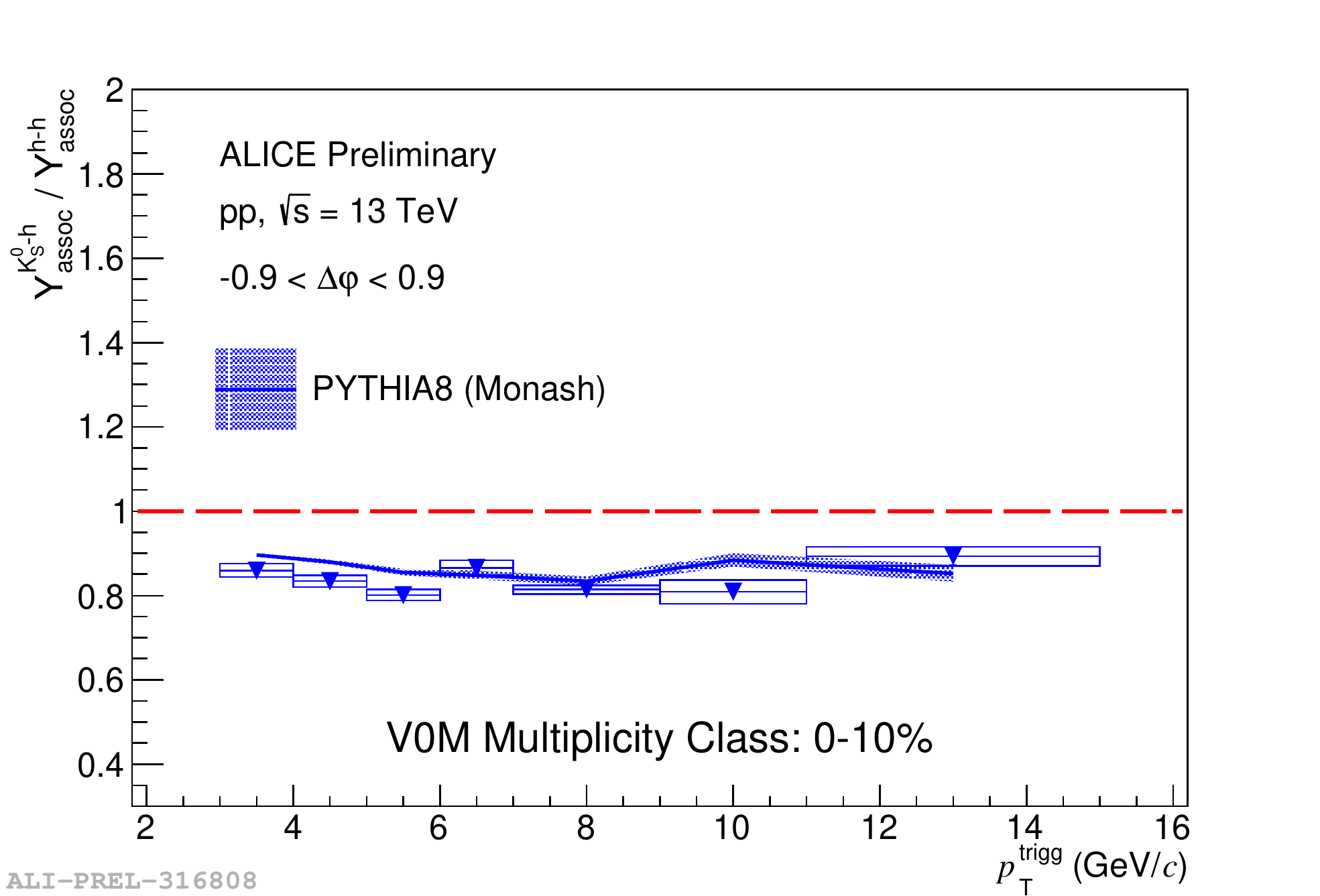}
\caption{Ratio of yields of ($\mathrm{K}^{0}_{\mathrm{S}}$-h/h-h) correlated pairs in the near-side jet peak compared to PYTHIA predictions ({\it shaded band}) for 0-10\% V0M multiplicity class pp events.}
\label{K0hyields}
\end{figure}

These yields are then compared to measurements from h-h correlations to look for differences in jet fragmentation. In all multiplicity classes, yields in both the near and away-side for $\mathrm{K}^{0}_{\mathrm{S}}$ triggers were lower than for unidentified hadron triggers across all associate $p_{\mathrm{T}}$ (fig. \ref{K0hyields}). However, simulated PYTHIA events have a ratio consistent with the measured data for both near and away-side yields.

\section{h-$\phi$ Angular Correlations}

\begin{figure}
\centering
\includegraphics[width = 3.9in]{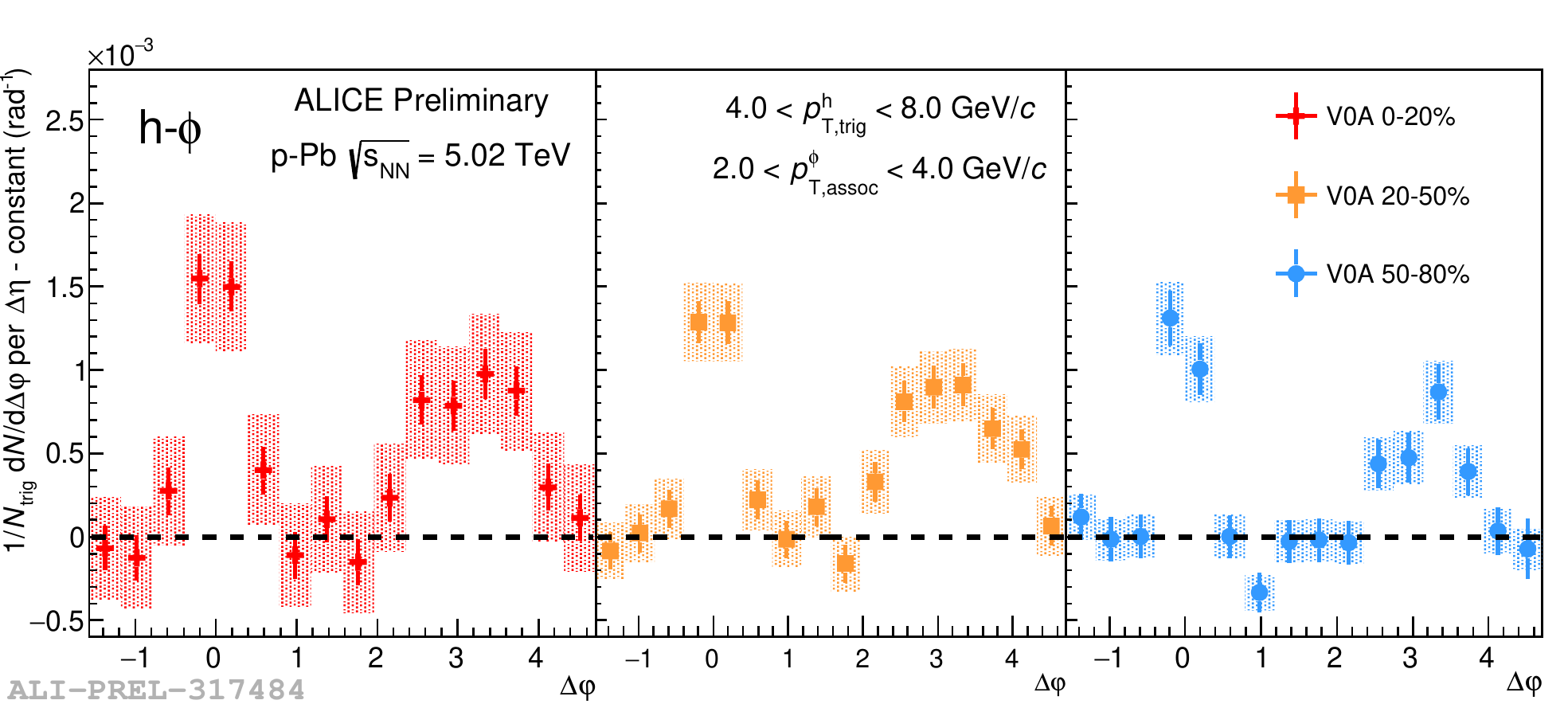}
\caption{Angular correlations between a high momentum hadron and intermediate momentum associated $\phi$ meson for different multiplicity classes.}
\label{hphicorr}
\end{figure}

Hadron-$\phi$ angular correlations are measured in p-Pb collisions at $\sqrt{s_{\mathrm{NN}}} = 5.02$ TeV to separate the $\phi$ produced within jets from $\phi$ produced in the underlying event. Candidate $\phi$'s are reconstructed from the $\phi \rightarrow \mathrm{K}^{+}\mathrm{K}^{-}$ decay channel (B.R. $\approx 49\%$). Trigger hadrons are selected with a transverse momentum of $4 < p^{\mathrm{h}}_{\mathrm{T}} < 8$ GeV/$c$ and correlated with ($\mathrm{K}^{+}\mathrm{K}^{-}$) pairs with momentum $2 < p^{\mathrm{KK}}_{\mathrm{T}} < 4$ GeV/$c$. Decay kaons are identified using PID cuts on signals from both the TPC and TOF detectors \cite{Adam:2016bpr}. The h-$\phi$ correlation is then obtained by subtracting combinatoric h-($\mathrm{K}^{+}\mathrm{K}^{-}$) correlated pairs (measured in the invariant mass sideband region) from the h-($\mathrm{K}^{+}\mathrm{K}^{-}$) correlation measured in the $\phi$ mass signal region. These correlations are measured for three different event multiplicity classes: 0-20\% (highest mult.), 20-50\%, and 50-80\% (fig. \ref{hphicorr}). Event multiplicity is determined using the A-side V0 detector (V0A) at large forward rapidity. Both the trigger hadron and associated $\phi$ are corrected for detector efficiency and acceptance effects.

After the h-$\phi$ correlation is measured, the yields of associated $\phi$ can be separated into a jet-like component and an underlying event component. To separate these, an assumed flat non-jet contribution is calculated as the average of the correlation in the regions away from the jet peaks. Once this underlying event is defined, the jet-like yields can be further separated into a near-side yield ($|\Delta\varphi| < \frac{\pi}{2}$) and an away-side yield ($\frac{\pi}{2} < \Delta\varphi < \frac{3\pi}{2}$). 

\begin{figure}
\centering
\includegraphics[width = 3.6in]{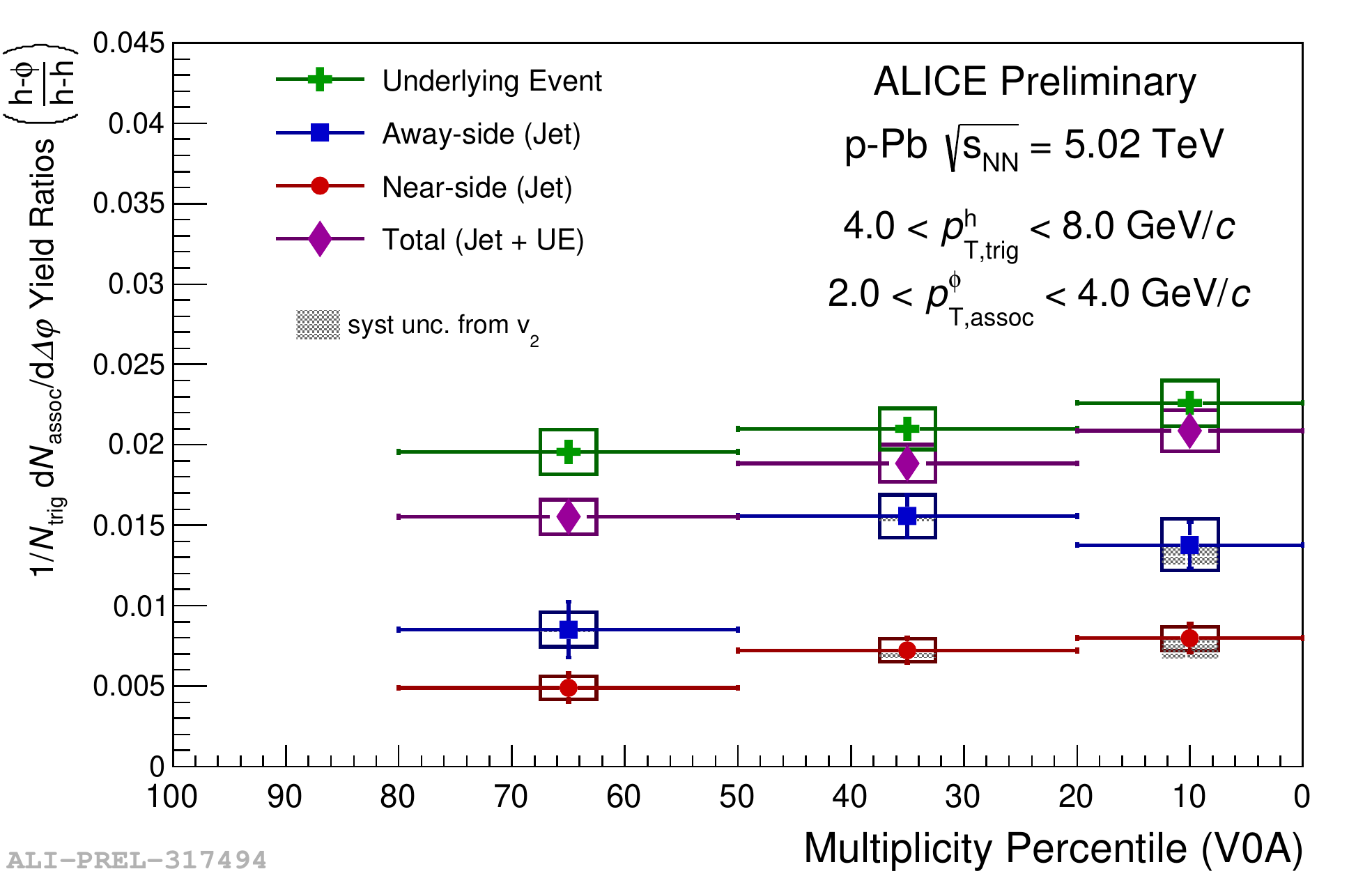}
\caption{Ratio of (h-$\phi$)/(h-h) pair yields in p-Pb events as a function of event multiplicity. Yield ratios are calculated both for inclusive total pairs, as well as for just the near and away-side jet regions and the underlying event region separately. Additional systematic uncertainties ({\it gray shaded boxes}) are included for the presence of a non-zero $v_2$ \cite{Abelev:2013aa}.}
\label{phiratio}
\end{figure}

Taking the ratio of h-$\phi$ pair yields to measured h-h pair yields gives a proxy measurement for $\phi$/h in jets and $\phi$/h in the underlying event (fig. \ref{phiratio}). The ratio of $\phi/$h in jets is seen to be systematically lower than the inclusive ratio, while the ratio in the underlying event is systematically higher. Further, as multiplicity increases so does the fraction of total particles coming from the underlying event. Together, these observations show that the rise of the inclusive ratio is in part due to higher multiplicity events being dominated by production in the underlying event regime.

\section{Conclusion}
The ALICE collaboration has measured $\mathrm{K}^{0}_{\mathrm{S}}$-h angular correlations as a function of multiplicity in pp events at $\sqrt{s} = 13$ TeV. This measurement compares the fragmentation of jets containing strangeness with inclusive charged jets. For all multiplicities, yields of associated hadrons with $\mathrm{K}^{0}_{\mathrm{S}}$ triggers were measured lower than predicted by simulated PYTHIA events. Taking the ratio to correlated pairs with an inclusive hadron trigger shows that $\mathrm{K}^{0}_{\mathrm{S}}$ triggers gave lower associated yields accross all multiplicities, but the ratio of ($\mathrm{K}^{0}_{\mathrm{S}}$-h)/(h-h) pairs is consistent with PYTHIA.

Hadron-$\phi$ angular correlations are measured as a function of multiplicity in p-Pb events at $\sqrt{s_{\mathrm{NN}}} = 5.02$ TeV. By measuring the yield of $\phi$ mesons associated with a high momentum trigger hadron, it is possibile to separately measure the ratio of $\phi$/h within jets and $\phi$/h in the underlying event. The $\phi$/h ratio in jets (hard production) is systematically lower than the inclusive ratio, while the $\phi$/h ratio in the underlying event (softer production) is systematically higher for all multiplicities.

\paragraph{Acknowledgements:}
This work was supported by U.S. Department of Energy Office of Science under contract number DE--SC0013391.
%
%

\end{document}